\documentclass[runningheads]{llncs}\usepackage{graphicx}
\usepackage{algorithm}  
\usepackage{algorithmicx}  
\usepackage{algpseudocode}  
\usepackage{amsmath} 
\usepackage{amsfonts}
\usepackage{booktabs}
\usepackage{multirow}
\usepackage{regexpatch}
\usepackage{amssymb}
\usepackage{mathrsfs}
\usepackage[misc]{ifsym} 
\usepackage[marginal]{footmisc}

\begin{document}

\title{Plug-and-play Shape Refinement Framework for Multi-site and Lifespan Brain Skull Stripping}
\titlerunning{Plug-and-play Shape Refinement Framework}
\author{Yunxiang Li\inst{1,2}, Ruilong Dan\inst{2}, Shuai Wang\inst{3}, Yifan Cao\inst{2}, Xiangde Luo\inst{4}, Chenghao Tan\inst{2}, Gangyong Jia\inst{2}, Huiyu Zhou\inst{5}, You Zhang\inst{1}, Yaqi Wang\inst{2,6}$^{(\textrm{\Letter})}$, Li Wang\inst{7}$^{(\textrm{\Letter})}$ }
\authorrunning{Y. Li et al.}
\institute{Department of Radiation Oncology, University of Texas Southwestern Medical Center, Dallas, USA \and Hangzhou Dianzi University, Hangzhou, China  \and School of Mechanical, Electrical and Information Engineering, Shandong University, Weihai, China  \and University of Electronic Science and Technology of China, Chengdu, China \and School of Computing and Mathematical Sciences, University of Leicester, UK \and Communication University of Zhejiang, Hangzhou, China \\  \email{wangyaqi@cuz.edu.cn} \and Developing Brain Computing Lab, Department of Radiology and BRIC,\\ University of North Carolina at Chapel Hill, Chapel Hill, USA \\  \email{li\_wang@med.unc.edu} }
\maketitle

\begin{abstract}
Skull stripping is a crucial prerequisite step in the analysis of brain magnetic resonance images (MRI). Although many excellent works or tools have been proposed, they suffer from low generalization capability. For instance, the model trained on a dataset with specific imaging parameters cannot be well applied to other datasets with different imaging parameters. Especially, for the lifespan datasets, the model trained on an adult dataset is not applicable to an infant dataset due to the large domain difference. To address this issue, numerous methods have been proposed, where domain adaptation based on feature alignment is the most common. Unfortunately, this method has some inherent shortcomings, which need to be retrained for each new domain and requires concurrent access to the input images of both domains. In this paper, we design a plug-and-play shape refinement (PSR) framework for multi-site and lifespan skull stripping. To deal with the domain shift between multi-site lifespan datasets, we take advantage of the brain shape prior, which is invariant to imaging parameters and ages. Experiments demonstrate that our framework can outperform the state-of-the-art methods on multi-site lifespan datasets.  
\end{abstract}
\footnote{Y. Li and R. Dan—Equal contribution.}

\section{Introduction}\label{sec:intro}
\begin{figure}[ht]
  \centering
  \includegraphics[width=0.85\textwidth]{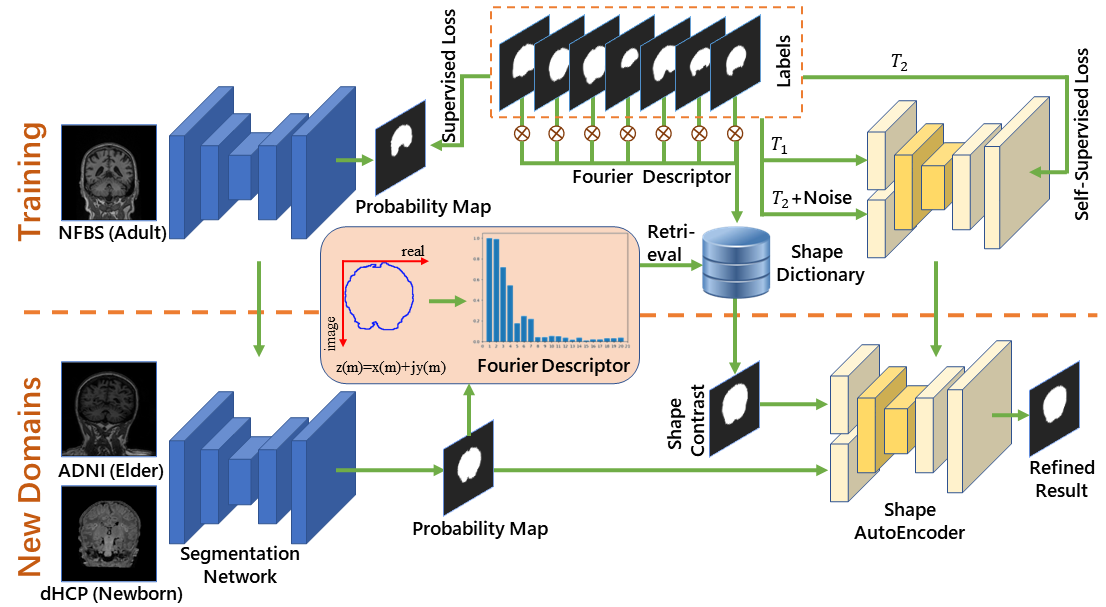}
  \caption{A schematic overview of our proposed plug-and-play shape refinement framework.}
  \label{fig:method}
\end{figure}

Skull stripping, the separation of brain tissue from non-brain tissue, is a critical preprocessing step for the characterization of brain MRI. Plenty of skull stripping tools have been proposed, e.g., morphology-based method: Brain Surface Extractor (BSE) \cite{shattuck2001magnetic} and surface-based method: Brain Extraction Tool (BET) \cite{smith2002fast}.
Compared with traditional skull stripping tools, deep learning has recently been proven more suitable for skull stripping, where 3D U-Net is the most popular backbone \cite{20163d,luo2021urpc}. Based on it, 3D Residual U-Net, 3D Attention UNet, and TransBTS are proposed to get better segmentation performance \cite{islam2019brain,yu2019liver,wang2021transbts}. 
In addition, some specific methods have been designed, e.g., Zhong et al. proposed a domain-invariant knowledge-guided attention network for brain skull stripping \cite{zhong2021dika}. Zhang et al. proposed a flattened residual network for infant MRI skull stripping \cite{zhang2019frnet}.
However, the high performance of deep learning-based methods requires that the training and testing datasets share a similar data distribution, which is hardly met due to a variety of device manufacturers, magnetic field strength, and acquisition protocols. Moreover, there is also a substantial data distribution difference across lifespan, e.g., the adult and infant brain MRI in Fig. \ref{fig:method}, where the infant's brain is undergoing myelination and maturation. 

Many efforts have been devoted to addressing the domain shift problem. Among them, the most widely used method is domain adaptation to align the latent feature distributions of the two domains \cite{li2021dispensed,dou2019pnp}. 
Unfortunately, there are some inherent limitations, including the need for retraining for each new domain and concurrent access to the input images of both domains. It is well known that medical images are difficult to share due to privacy. Compared with the original image, manual labels contain much less privacy information and are easier to share. Therefore, we propose a plug-and-play shape refinement (PSR) framework in this work. The main contributions of our method are three-fold: 
\textbf{1)} A novel plug-and-play segmentation result refinement framework is designed.
\textbf{2)} A shape dictionary based on the Fourier Descriptors \cite{NIXON2012343} is proposed to fully utilize the anatomical prior knowledge of the brain shape.
\textbf{3)} To better model the overall shape information, we designed a shape AutoEncoder (SAE) based on Shuffle Transformer.

\section{Method}
\subsection{Overall Architecture}
The overall framework of PSR is illustrated in Fig. \ref{fig:method}. Specifically, we can arbitrarily take a model trained in the source domain as the segmentation network. Fourier Descriptors \cite{NIXON2012343} of the source labels are computed to build a shape dictionary.
Then, the new domain image is directly input into the model, and Fourier Descriptors of the segmentation results are computed. Through it, we can retrieve a label with the closest shape from the shape dictionary. 
Finally, the segmentation results, together with the retrieved labels, serve as inputs into SAE to further refine the segmentation results.

\subsection{Shape Dictionary}
To make full use of the anatomy prior knowledge of brain shape, we calculate the corresponding Fourier Descriptors for each subject according to the source labels and store them in the dictionary. Assuming source labels $L^{s}=\{ {l_{i}^{s}} \}_{i=1}^{\left| L^{s} \right|}$, the constructing process of shape dictionary $D^{s}$ is defined as Eq. (\ref{shape_dict_construct})

\begin{equation}\label{shape_dict_construct}
    D^{s} = \{ d_{i}^{s} \mid d_{i}^{s}=F({l_{i}^{s}}) \}_{i=1}^{\left| L^{s} \right|}
\end{equation}

\noindent where $F$ is the Fourier Descriptor for a quantitative representation of closed shapes independent of their starting point, scale, location, and rotation. 
The whole process of computing Fourier Descriptors consists of three steps: (1) Establish a coordinate system in the upper left corner of the boundary, and the coordinate axis is tangent to the boundary.
(2) Take the two axes as real and imaginary numbers respectively, and the coordinates of points $(x_m,y_m)$ on the boundary are expressed in complex numbers $z(m) = x_m + j y_m$.
(3) The discrete Fourier transform (DFT) is applied to the above coordinates to obtain the Fourier Descriptor of the boundary shape, which is defined as $Z(k)$ in Eq. (\ref{DFT}). 
\begin{equation}\label{DFT}
    Z(k) = \frac{1}{N}\sum_{m=0}^{N-1} z(m)e^{-j2\pi mk /N},k = 0, 1, 2, …, N-1
\end{equation}
\noindent where $N$ is the amount of the boundary points. Specifically, we choose 10 low-frequency coefficients as the final Fourier Descriptors, which is sufficient to achieve the required accuracy for retrieving \cite{dalitz2013fourier}. 

\begin{figure}[ht]
  \centering
  \includegraphics[width=0.85\textwidth]{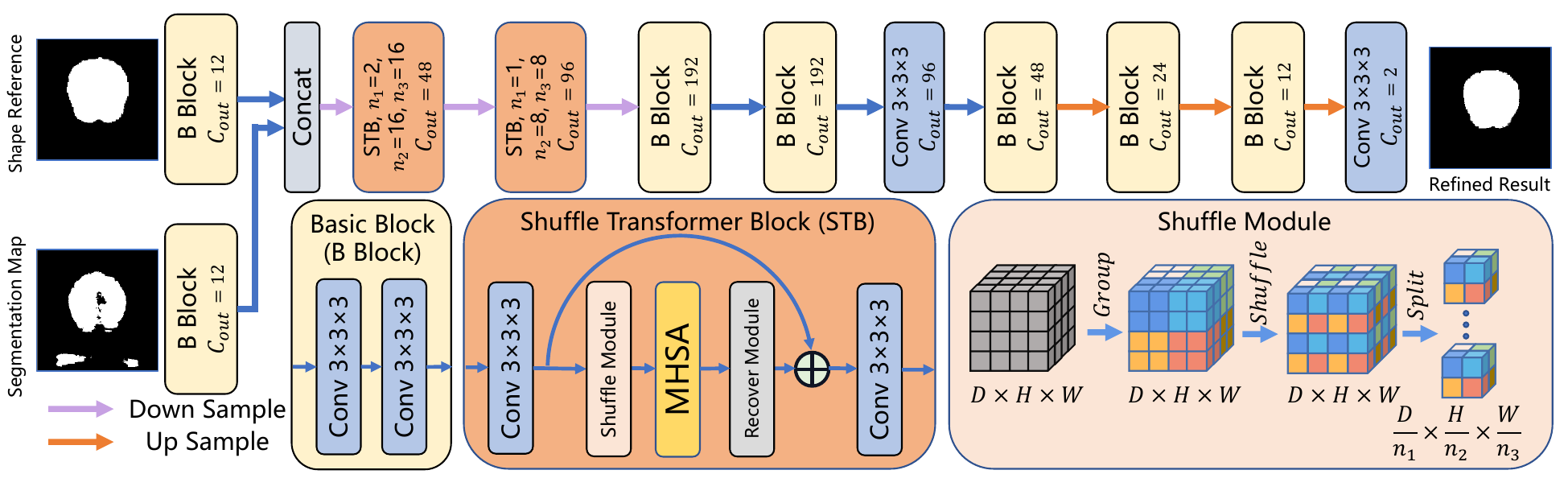}
  \caption{A schematic overview of our Shape AutoEncoder. The skip connection of the U-shaped structure is not shown for simplicity.}
  \label{fig:sta}
\end{figure}

\subsection{Shape AutoEncoder (SAE)}
The Shape AutoEncoder is designed to further refine the segmentation results based on both segmentation maps from the segmentation network and the shape reference from the shape dictionary, with the detailed structure shown in Fig. \ref{fig:sta}.  
To better extract global information such as overall shape, Transformer-based methods are good candidates \cite{liu2021swin,li2021agmb,li2021gt}. Based on 2D Shuffle Transformer \cite{huang2021shuffle}, we design a 3D Shuffle Transformer to capture global capability.

\subsubsection{Shuffle Transformer:}
Our shuffle Transformer mixes the voxel features regularly, then evenly divides them into non-overlapping groups. Specifically, Shuffle Module takes the feature $X$ as input and outputs shuffled blocks $X^{b}$; thus it can be formulated through Eq. (\ref{S_S_G_op}).


\begin{equation} \label{S_S_G_op}
    X^{b} = {\{ x_{i}^{b} \mid x_{i}^{b}=Split(Shuffle(Group(X, i))), x_{i}^{b} \in \mathbb{R}^{ \frac{D}{n_{1}} \times \frac{H}{n_{2}} \times \frac{W}{n_{3}} } \} }_{i=1}^{(n_{1} \times n_{2} \times n_{3})}
\end{equation}
where the details of Group, Shuffle, and Split are depicted in Fig. \ref{fig:sta}. After these operations we obtain a total number of $(n_1 \times n_2 \times n_3)$ shuffled blocks $x^b$, each with a size of $\frac{D}{n_1} \times \frac{H}{n_2} \times \frac{W}{n_3}$. Subsequently, each group is computed by multi-head self-attention. Since our shuffle operation keeps the relative position of each element in the shuffled blocks as the same as the original feature block, our position encoding method is based on relative-distance-aware position encoding \cite{ramachandran2019stand,Bello_2019_ICCV}. The query-key-value (QKV) attention \cite{vaswani2017attention} in each small block $x^b$ can be computed by Eq. (\ref{self_atten}).
 \begin{equation} \label{self_atten}
     Attention(Q,K,V) = softmax(\frac{QK^{T}}{\sqrt{d_{k}}} + B)V
 \end{equation}
where $Q$, $K$, and $V$ stand the query, key, and value matrices of dimension $d_{k}$ respectively, and $B$ denotes the position biases matrix.

\subsubsection{Self-supervised Training of Shape AutoEncoder:}
As for the training process of SAE, it is based on self-supervision, meaning the supervisory signals are generated by the input, and we use the shared source labels as the training images of SAE. Each image is processed by two kinds of random spatial transformations ($T_{1}$ and $T_{2}$). Specifically, our random spatial transformations contain random scaling, random rotation, and random clipping.
Assume we have images $Y={\{y_{i}\}}_{i=1}^{\eta}$ where $\eta$ denotes the number of images, thus transformed images are generated via Eq. (\ref{trans_func}).

\begin{equation} \label{trans_func}
\begin{aligned}
    &Y^{T_{1}}&=\{y_{i}^{T_{1}} \mid y_{i}^{T_{1}}=T_{1} (y_{i})\}_{i=1}^{\eta} \\
    &Y^{T_{2}}&=\{y_{i}^{T_{2}} \mid y_{i}^{T_{2}}=T_{2} (y_{i})\}_{i=1}^{\eta} 
\end{aligned}
\end{equation}
To mimic the unreliable segmentation results caused by domain shift, random noises $RN$ are added to $T_{2}$ through placing random false positive and false negative stain to generate $Y^{{T_{2}} \circ {RN}}$ formulated by Eq. (\ref{gen_t2_noise}). The random noise is controlled by two variables: the amount and size of the noise. The hyperparameters of the random noise are discussed in the supplementary material. 
\begin{equation} \label{gen_t2_noise}
    Y^{{T_{2}} \circ {RN}}={RN(Y^{T_{2}})}
\end{equation}
Let ${SAE}{(y_{i}^{T1}, y_{i}^{{T_{2}} \circ {RN}};\theta)}$ be the Shape AutoEncoder, thus refined outputs $\hat{Y}$ are defined via Eq. (\ref{sta_proc}).

\begin{equation}\label{sta_proc}
    \hat{Y} = {SAE}{(Y^{T_{1}}, Y^{{T_{2}} \circ {RN}};\theta)}
\end{equation}
$\hat{Y}$ and $\theta$ represent the prediction output and the trainable parameters of SAE respectively, and the self-supervised loss function of the SAE is defined by Eq. (\ref{sae_loss}).

\begin{equation}\label{sae_loss}
\begin{aligned}
    \mathcal{L}_{SAE}(\hat{Y},Y^{T_2})&= -\frac{1}{\eta}\sum_{i=1}^{\eta} (T_2(y_{i}) \times ln(SAE(T_1(y_i),RN(T_2(y_i));\theta)) \\
    &+ (1-T_2(y_{i})) \times (1-ln(SAE(T_1(y_i),RN(T_2(y_i));\theta)))) 
\end{aligned}
\end{equation}
In this way, it can not only combine segmentation results and shape reference but also denoise and refine the unreliable segmentation results automatically through learning shape prior knowledge from the labels.

\section{Implementation and Experiments}
\subsection{Datasets and Evaluation Metrics} 
We evaluated the proposed method on the publicly available dataset, where the source domain is from Neurofeedback Skull-stripped (NFBS) \cite{eskildsen2012beast}, and the new (target) domains are from Alzheimer’s Disease Neuroimaging Initiative (ADNI) \cite{jack2008alzheimer} and Developing Human Connectome Project (dHCP) \cite{makropoulos2018developing}. Note that subjects from NFDS are young adults from 21 to 45 years old and ADNI are older adults from 55 and 90 years old, and dHCP are newborns. After resampling and padding, the size of the individual scan is $256 \times 256 \times 256$ and each voxel size is $1 \times 1 \times 1 \ mm^3$.
We selected 25 subjects from NFBS with manual labels as the training dataset and each 10 subjects from ADNI and dHCP as the testing dataset, and 3-fold cross-validation is used. It is worth noting that there is no available publicly manual label of dHCP, and thus we only compare the results by visual inspection. Due to the limited GPU memory, a sub-volume of size $64\times 64\times 64$ is used as the first stage segmentation network input. In order to better capture the overall shape, the input size of the Shape AutoEncoder is set to $8 \times 256 \times 256$, and the middle layer slice with the size of $256 \times 256$ is used to compute the Fourier Descriptors and retrieve the most similar shape from the dictionary. For all experiments described below, the Average Surface Distance (ASD), the Dice Coefficients (DICE), the sensitivity (SEN) and specificity (SPE) are chosen for evaluation metrics.

\subsection{Implementation Details}
All our experiments were based on the PyTorch framework and carried out on 4 Nvidia RTX 2080Ti GPUs. We trained our network from scratch for a total of 10000 iterators, and the parameters were updated by the Adam algorithm (momentum = 0.97, weight decay = $5 \times 10^{-4}$). We adopt a batch size of 4 and set the learning rate as $2 \times 10^{-4}$. Notedly, if there is no additional statement, our PSR is combined with 3D U-Net by default in this paper. The discussion of the hyperparameters is on the supplementary material.

\begin{table}[ht]
\caption{Comparison with the state-of-the-art methods on ADNI.}
  \label{tab:adni}
  \centering
  \setlength{\tabcolsep}{0.35mm}
\begin{tabular}{c|c|c|c|c}
\hline
Method  & ASD (mm) & DICE (\%) & SPE (\%) & SEN (\%)   \\ \hline
3D U-Net \cite{20163d} & 11.57±9.83& 88.30±3.53 & 99.51±0.68 & 82.69±3.10 \\ 
CycleGAN \cite{zhu2017unpaired} & 18.96±7.88 & 86.27±2.85& 98.72±0.51&85.19±2.58\\
EMNet \cite{sun2021multi} &  8.67±9.86  &  91.49±2.39 & 99.37±0.49 & \textbf{89.32±2.10} \\
Tent \cite{wang2021tent} & 7.19±6.24  &  90.52±1.70 & \textbf{99.64±0.28} & 85.60±2.18 \\
\textbf{PSR (3D U-Net)}  &\textbf{4.63±0.98} &\textbf{91.57±1.38}& 99.55±0.15 &88.09±2.67 \\
\hline
\end{tabular}
\end{table}

\begin{figure}[ht]
  \centering
  \includegraphics[width=0.85\textwidth]{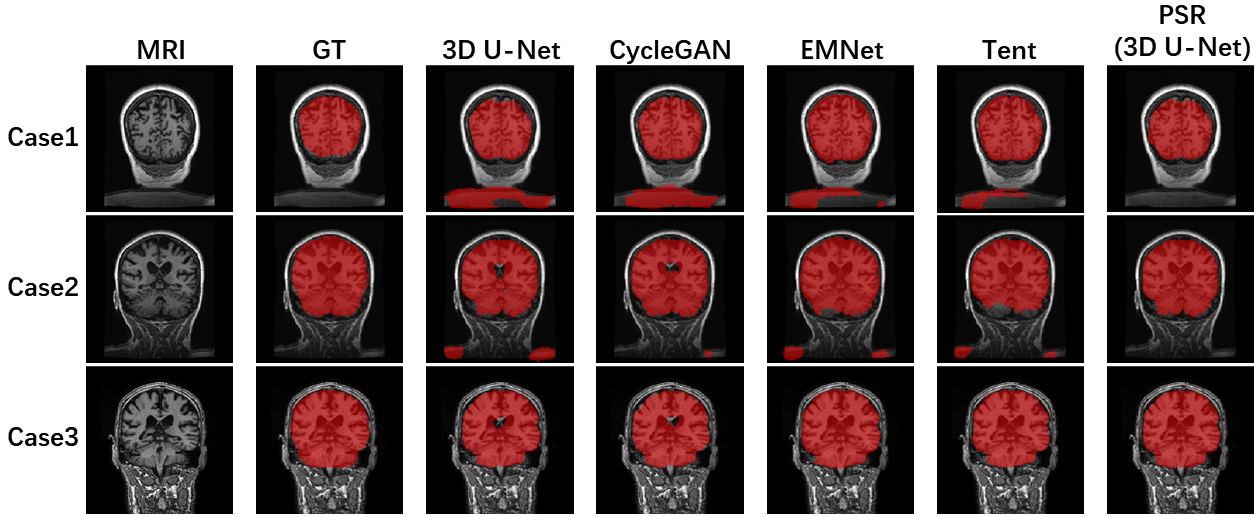}
  \caption{Visualization of segmentation results on ADNI.}
  \label{fig:adni}
\end{figure}

\subsection{Experimental Results on Cross-site Dataset}
Our results are presented in Table. \ref{tab:adni}, and our method is combined with 3D U-Net, namely PSR (3D U-Net).
We can observe that our method enhances the performance of 3D U-Net a lot.
Moreover, our method outperforms the state-of-the-art domain adaptation method, i.e., EMNet. 
Interestingly, the ASD achieved by other methods was unexpectedly far higher than ours. This may be due to the fact that the training data from the source domain do not have the shoulder part, but the testing data in the new domain do have the shoulder, which is illustrated in Fig. \ref{fig:adni}. Our method is capable of identifying the segmentation result of the shoulder as unreliable results and excluding it from the brain tissues.
We can also notice that the output of CycleGAN mistakenly identifies many non-brain regions as the brain. Consequently, its performance is unexpectedly poorer than those obtained from the source model without domain adaptation.

In order to further test the potential of our PSR, we also combined it with other popular networks. As shown in Table. \ref{tab:combined}, the segmentation performances of all the networks have been enhanced a lot by combining with our PSR.

\begin{table}[ht]
\caption{Comparison with the state-of-the-art methods on ADNI.}
  \label{tab:combined}
  \centering
  \setlength{\tabcolsep}{3.65mm}
\begin{tabular}{c|c|c}
\hline
Method  & ASD (mm) & DICE (\%)  \\ \hline
3D Residual U-Net  & 13.73±9.69& 88.99±4.19  \\
\textbf{PSR (3D Residual U-Net)}  &\textbf{4.42±1.61} &\textbf{91.34±1.61}\\ \hline
3D Attention U-Net  & 10.13±7.03& 89.29±2.17  \\
\textbf{PSR (3D Attention U-Net)}  &\textbf{4.06±0.76} &\textbf{91.22±0.99}\\ \hline
TransBTS  & 12.31±12.26& 88.80±4.67  \\ 
\textbf{PSR (TransBTS)}  &\textbf{6.01±2.64} &\textbf{90.87±1.65}\\ 
\hline
3D U-Net & 11.57±9.83& 88.30±3.53  \\ 
\textbf{PSR (3D U-Net)}  &\textbf{4.63±0.98} &\textbf{91.57±1.38} \\
\hline
\end{tabular}
\end{table}

\subsection{Visualization Results on Newborns }
\begin{figure}[ht]
  \centering
  \includegraphics[width=1\textwidth]{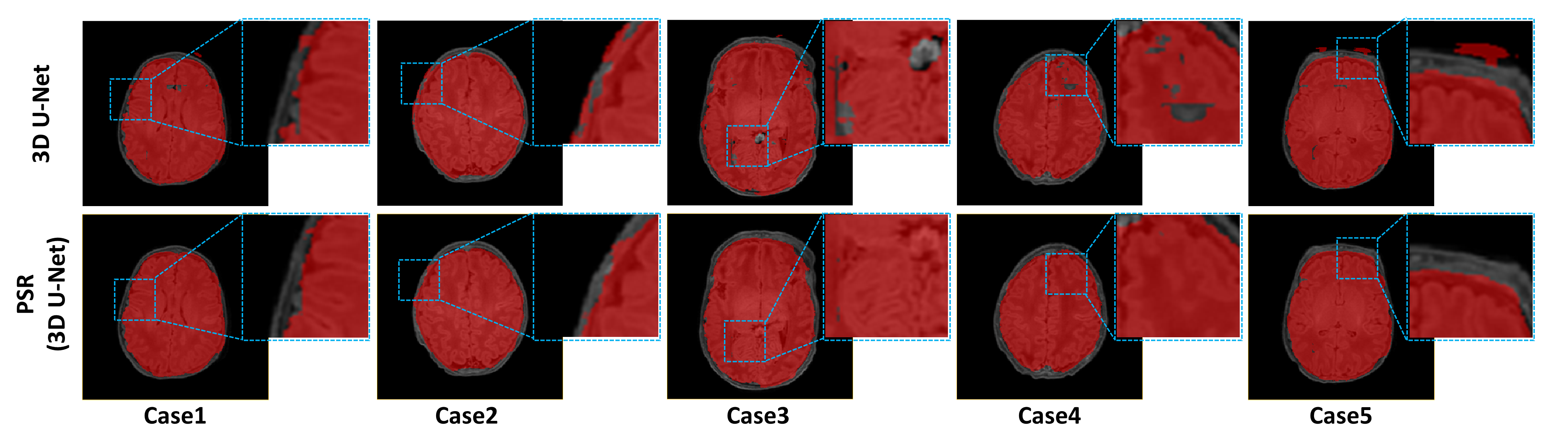}
  \caption{Qualitative comparison of segmentation results on newborn brain MRI from dHCP. }
  \label{fig:dhcp}
\end{figure}

We qualitatively compare the segmentation outputs of the proposed PSR and the 3D U-Net on the newborn. Although the infant shares a similar shape and basic structure with the adult, the infant's brain develops rapidly throughout the first year of life, resulting in huge appearance differences from the adult. As shown in Fig. \ref{fig:dhcp}, 3D U-Net cannot achieve accurate brain tissues with fuzzy brain boundaries, while our method is able to generate smooth and reasonable segmentations.

\subsection{Ablation Study}
In order to verify the effectiveness of the main components of our proposed PSR, we conducted the following ablation study: a) implement our PSR without Shape AutoEncoder, referred to as PSR w/o SAE. b) implement our PSR without shape dictionary, that is, only the unrefined segmentation results are input to Shape AutoEncoder, referred to as PSR w/o SD. c) implement our PSR without shuffle Transformer, that is, replace the shuffle Transformer block with the basic convolution block, denoted as PSR w/o ST. 
Quantitative comparison results of the three variants along with the PSR are illustrated in Table \ref{tab:ablation}. The proposed PSR achieves improved performance, especially in terms of important ASD and DICE metrics.

\begin{table}[ht]
\caption{Ablation study of our method.}
  \label{tab:ablation}
  \centering
  \setlength{\tabcolsep}{2mm}
\begin{tabular}{c|c|c|c|c}
\hline
Method & ASD (mm) & DICE (\%)  & SPE (\%) & SEN (\%)   \\ \hline
PSR w/o SAE  & 11.57±9.83 & 88.30±3.53 & 99.51±0.68 & 82.69±3.10\\
PSR w/o SD & 6.86±3.53 & 89.73±1.42 & 99.41±0.13 & 86.04±2.21 \\ 
PSR w/o ST  & 6.39±4.24 & 90.62±1.40 & \textbf{99.76±0.13} & 84.79±2.66  \\ 
\textbf{PSR} &\textbf{4.63±0.98} &\textbf{91.57±1.38} & 99.55±0.15 &\textbf{88.09±.2.67} \\
\hline
\end{tabular}
\end{table}

\section{Conclusion}
In summary, we presented a plug-and-play shape refinement framework and successfully applied it to the skull stripping task on multi-site lifespan datasets. Our method consists of a shape dictionary and a Shape AutoEncoder. 
With the assistance of the shape dictionary, the Shape AutoEncoder makes full use of the anatomical prior knowledge to refine the unreliable segmentation results in the new domain.
Experimental results demonstrated that the proposed method can enhance the performance of most networks and achieves better performance than the state-of-the-art unsupervised domain adaptation methods, and the proposed Shape AutoEncoder can further enhance traditional skull stripping tools. Theoretically, our method can also be widely applied to other organ segmentation.\\\\
\textbf{Acknowledgements.} This work was supported in part by National Institutes of Health (Grant No. R01CA240808 and R01CA258987), National Natural Science Foundation of China (Grant No. U20A20386), Shandong Provincial Natural Science Foundation (Grant No. 2022HWYQ-041)

\bibliographystyle{unsrt}
\begingroup
  
  \small 
  \bibliography{paper}
\endgroup

\end{document}